\documentclass[conference]{IEEEtran}
\IEEEoverridecommandlockouts
\usepackage{cite}
\usepackage{amsmath,amssymb,amsfonts}
\usepackage{algorithmic}
\usepackage{graphicx}
\usepackage{textcomp}
\usepackage{xcolor}
\usepackage{booktabs}
\usepackage{hyperref}
\usepackage{graphicx,url,times,booktabs,tabularx}
\usepackage{bm}
\usepackage{setspace}
\usepackage{comment}
\usepackage{enumitem}
\def\BibTeX{{\rm B\kern-.05em{\sc i\kern-.025em b}\kern-.08em
    T\kern-.1667em\lower.7ex\hbox{E}\kern-.125emX}}

\usepackage{contour}
\usepackage[normalem]{ulem}

\contourlength{0.7pt}

\newcommand{\myuline}[1]{%
  \uline{\phantom{#1}}%
  \llap{\contour{white}{#1}}%
}

\begin{document}

\title{Noisy-target Training: A Training Strategy for DNN-based Speech Enhancement \\ without Clean Speech}

\author{Takuya Fujimura$^{\dagger}$, Yuma Koizumi$^{\ddag}$, Kohei Yatabe$^{\star}$, Ryoichi Miyazaki$^{\dagger}$\\
%$^{\ddag}$NTT Corporation, Tokyo, Japan\qquad\qquad
%$^{\star}$Waseda University, Tokyo, Japan
\textit{$^{\dagger}$National Institute of Technology, Tokuyama College, Yamaguchi, Japan} \\
\texttt{\{%
  \href{mailto:i15fujimura1t@tokuyama.kosen-ac.jp}{i15fujimura1t},%
  \href{mailto:miyazaki@tokuyama.kosen-ac.jp}{miyazaki}%
  \}@tokuyama.kosen-ac.jp} \\
\textit{$^{\ddag}$NTT Corporation, Tokyo, Japan} \qquad\qquad
\textit{$^{\star}$Waseda University, Tokyo, Japan}\\
\texttt{%
  \href{mailto:koizumi.yuma@ieee.org}{koizumi.yuma}%
  @ieee.org.jp} \qquad\qquad
\texttt{%
  \href{mailto:k.yatabe@asagi.waseda.jp}{k.yatabe}%
  @asagi.waseda.jp}
%\vspace{-10pt}
}

\maketitle

\begin{abstract}
Deep neural network (DNN)--based speech enhancement ordinarily requires clean speech signals as the training target.
However, collecting clean signals is very costly because they must be recorded in a studio.
This requirement currently restricts the amount of training data for speech enhancement to less than 1/1000 of that of speech recognition which does not need clean signals.
Increasing the amount of training data is important for improving the performance, and hence the requirement of clean signals should be relaxed.
In this paper, we propose a training strategy that does not require clean signals.
The proposed method only utilizes noisy signals for training, which enables us to use a variety of speech signals in the wild.
Our experimental results showed that the proposed method can achieve the performance similar to that of a DNN trained with clean signals.
\end{abstract}

\begin{IEEEkeywords}
Single-channel speech enhancement, deep neural network (DNN), training target, Noise2Noise.
\end{IEEEkeywords}

%%%%%%%%%%%%%%%%%%%%%%%%%%%%%%%%%%%%%%%%%%%%%%%%%%%%%%%%
\section{Introduction}
\label{sec:intro}
%%%%%%%%%%%%%%%%%%%%%%%%%%%%%%%%%%%%%%%%%%%%%%%%%%%%%%%%

%%%%%%%%%%%%%%%%%%%
% DNN-based speech enhancement の論文
% A. Narayanan and D. Wang, “Ideal ratio mask estimation using deep neural networks for robust speech recognition,” in Proc. ICASSP, 2013.
% Y.Xu,J.Du,L.R.Dai,andC.H.Lee,“An experimental study on speech enhancement based on deep neural networks,” IEEE Signal Processing Letters, pp.65–68, 2014.
% H. Erdogan, J. R. Hershey, S. Watanabe, and J. Le Roux, “Phase-Sensitive and Recognition-Boosted Speech Separation using Deep Recurrent Neural Networks,” Proc. of Int. Conf. on Acoust., Speech, and Signal Process. (ICASSP), 2015.
% D. S. Williamson, Y. Wang and D. L. Wang, “Complex Ratio Masking for Monaural Speech Separation,” IEEE/ACM Trans. on Audio, Speech, and Lang. Process., 2016.
% S. Pascual, A. Bonafonte, and J. Serra, “SEGAN: Speech En- hancement Generative Adversarial Network,” Proc. of Inter- speech, 2017.
% S.W.Fu,C.F.Liao,Y.Tsao,andS.D.Lin,“MetricGAN:Gen- erative Adversarial Networks based Black-box Metric Scores Optimization for Speech Enhancement,” Proc. of Int. Conf. on Machine Learning (ICML), 2019.
%%%%%%%%%%%%%%%%%%%

\begin{figure}[t]
\vspace{0pt}
  \centerline{\includegraphics[bb=0 0 539   282,width=\linewidth]{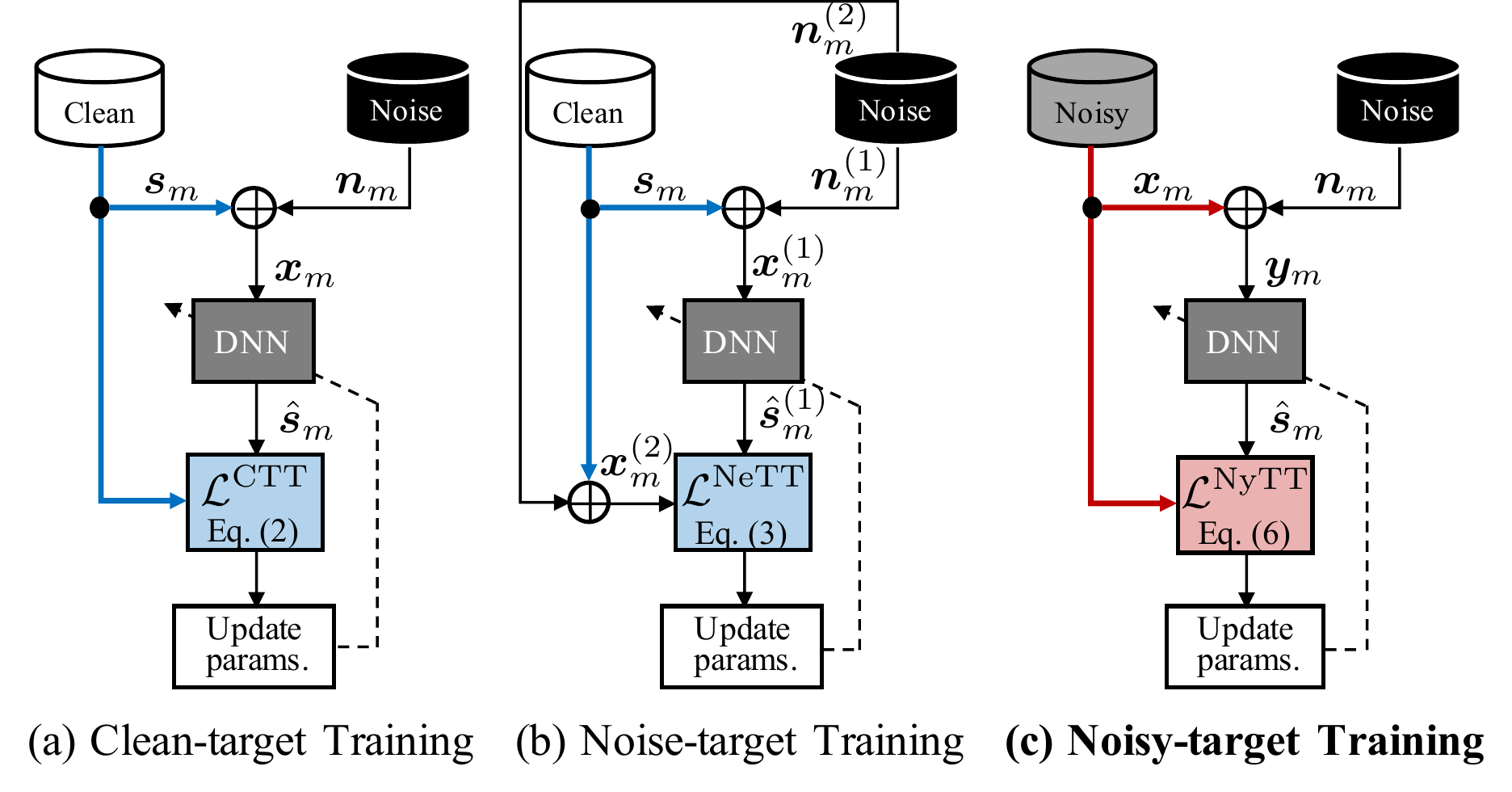}}
  %\vspace{-10pt}
  \caption{
  Overview of
  (a) Clean-target Training,
  (b) Noise-target Training, and
  (c) Noisy-target Training (porposed method). The proposed Noisy-target Training does not use any clean speech during training, in contrast to conventional Clean-target and Noise-target Training.
  }
\label{fig:methods}
  \vspace{-10pt}
\end{figure}

Speech enhancement is utilized for recovering target speech from a noisy observed signal~\cite{Wang_2018}.
It is a fundamental task with a wide range of applications, including automatic speech recognition (ASR)~\cite{Erdogan_2015}.
%\cite{NTTchime,narayanan_2013,Erdogan_2015}.
%Over the last decade, a rapid progress has been made by using supervised training of deep neural networks~(DNN)~\cite{Wang_2018, Koizumi_ICASSP_2017, Koizumi_TASL_2018, Koizumi_icassp_2020, Takeuchi_icassp_2019, Takeuchi_icassp_2020_Real_time, Takeuchi_icassp_2020_Invertible, kawanaka}.
Over the last decade, a rapid progress has been made by using supervised training of deep neural networks~(DNN)~\cite{Erdogan_2015, Koizumi_icassp_2020, kawanaka}.
%\cite{ narayanan_2013, Erdogan_2015, Xu_2014, williamson_2016, Pascual_2017, Fu_2019, Koizumi_TASL_2018, Koizumi_icassp_2020, Takeuchi_icassp_2019, Takeuchi_icassp_2020_Real_time, kawanaka}.
A DNN is trained so that it predicts the target speech from an input noisy observation. 
In the training, the \myuline{training target is clean speech}, and the input signal is simulated by \myuline{using clean speech and noise} as shown in Fig.\:\ref{fig:methods}\,(a).
In this paper, we refer to this standard training strategy as \textbf{Clean-target Training} (CTT).

Although Clean-target Training is clearly a proper strategy, it has two potential problems.
First, collecting studio-recorded signals is very costly and time-consuming.
Unlike image, speech signals are easily contaminated due to the surrounded environment.
Thus, ``clean'' signals can only be acquired under well-controlled conditions in a studio.
Such difficulty prohibits collecting a huge amount of data for training.
In fact, a typical dataset in speech enhancement contains only 12 thousand utterances~\cite{voicebankdemand}, whereas training of an ASR system may utilize over 35 million utterances~\cite{He_2019} because ASR does not require clean speech as the target.
Second, providing enough variations of the recording condition for the training is hopeless.
The real-world observations vary with multiple factors, including recording equipment, mouth-microphone distance, and the Lombard effect.
Simulating all of these factors in a studio to obtain a variety of clean signals is impossible.
Hence, the training dataset and real-world data have mismatches that can degrade the performance of speech enhancement, e.g., a training dataset is recorded with a studio (large-diaphragm condenser) microphone, while a real-world signal is recorded with a (low-priced MEMS) microphone implemented in a smartphone.
Because of these reasons, using clean speech signals as the training target can be an essential limitation.

% T. Ochiai, S. Watanabe, T. Hori, and J. R. Hershey, “Multichannel End-to-end Speech Recognition,” in Proc. ICML, 2017.
% Scott Wisdom, Efthymios Tzinis, Hakan Erdogan, Ron J. Weiss, Kevin Wilson, John R. Hershey "Unsupervised Sound Separation Using Mixtures of Mixtures," ICML 2020 Workshop on Self-supervision in Audio and Speech

To overcome such limitation, some research has attempted to train a DNN without the clean target~\cite{scott_2020, n2n_speech}.
%\cite{ochiai_2017,scott_2020, n2n_speech}.
An interesting strategy aiming at this goal is \textbf{Noise-target Training} (NeTT) as shown in Fig.\:\ref{fig:methods}\,(b).
Its \myuline{training target is mixture} \myuline{of speech and noise}, and the input signal is simulated by \myuline{using the same clean speech and some other noise}.
This training strategy was originally proposed in image processing by the name Noise2Noise~\cite{noise2noise}, and was later applied to speech enhancement~\cite{n2n_speech}.
In Noise2Noise, pairs of noisy signals that consist of different noises and \textit{exactly} the same clean target are utilized for training.
Since the contained clean signal is the same, the trained DNN tries to map the noise in an input signal to another noise.
Assuming that the noise distribution is zero-mean, the trained DNN becomes a noise suppressor \cite{noise2noise}.
By Noise2Noise training, the requirement of a clean target can be avoided for image applications because photos of exactly the same target with different noise can be easily obtained by a camera with multiple exposures.
However, it is not applicable to audio applications because it is impossible to observe multiple noisy signals with exactly the same speech signal.
Therefore, the previous research has utilized clean speech to simulate a pair of noisy signals \cite{n2n_speech} as in Fig.\:\ref{fig:methods}\,(b), which inherits the limitation of Clean-target Training.

In this paper, we propose \textbf{Noisy-target Training} (NyTT) as in Fig.\:\ref{fig:methods}\,(c).
Its \myuline{training target is noisy speech}, and the input signal is simulated by \myuline{using the same noisy speech and noise}.
In other words, a DNN is trained to predict a noisy speech signal from a more noisy signal.
Although this strategy might sound inappropriate, it can achieve results similar to those obtained by Clean-target Training when there is a mismatch between training and testing datasets.
Our contributions are as follows: (1) proposing a new training strategy, (2) examining several training conditions by extensive experiment, and (3) analyzing the experimental results.

% In this paper, we propose a new training strategy called {\it w/o-Clean-Training}, which does not use any clean speech while training. This strategy use [more-noisy]-noisy pair which is synthesized by mixing a noisy speech and a noise. Then, the DNN predicts clean speech from more-noisy signal, and is trained to minimize the prediction error between the noisy speech and output as shown in Fig.\,\ref{fig:methods}\,(c).
% This strategy allows to use the large number of noisy samples recorded in various real environments, as well as it can be expected to overcome the training/testing domain mismatch problem.
% Experimental results show that (i) the proposed method can perform speech enhancement without clean, and (ii) outperform Clean-Training on the mismatched domain.

%\vspace{-4pt}
\section{Related works}
\label{sec:related}
%\vspace{-4pt}

% \subsection{DNN-based Speech Enhancement}
Let $T$-point-long time-domain observation $\bm{x} \in \mathbb{R}^T$ be a mixture of a target speech $\bm{s}$ and observation noise $\bm{n}^{(\mathrm{obs})}$ as $\bm{x} = \bm{s} + \bm{n}^{(\mathrm{obs})}$. 
The goal of speech enhancement is to recover $\bm{s}$ only from $\bm{x}$.
Over the last decade, application of DNN to speech enhancement has substantially advanced the state-of-the art performance~\cite{Wang_2018, Erdogan_2015, Koizumi_icassp_2020, kawanaka}.
%\cite{Wang_2018, narayanan_2013, Erdogan_2015, Xu_2014, williamson_2016, Pascual_2017, Fu_2019, Koizumi_TASL_2018, Koizumi_icassp_2020, Takeuchi_icassp_2019, Takeuchi_icassp_2020_Real_time, kawanaka}.

A popular method is to utilize a DNN for estimating a time-frequency (T-F) mask in the short-time Fourier transform (STFT)--domain~\cite{Wang_2018}. 
Let $\mathcal{F}: \mathbb{R}^T\! \to \mathbb{C}^{F \times K}$ denote STFT, where $F$ and $K$ are the numbers of frequency and time bins, respectively.
Then, DNN-based speech enhancement can be written as
%\vspace{-2pt}
\begin{equation}
\hat{\bm{s}} = 
\mathcal{F}^{\dagger} \left(
\mathcal{M}( \bm{x}; \theta ) \odot 
\mathcal{F} \left( \bm{x}
\right)
\right),
\label{eq:DNNusage}
%\vspace{-2pt}
\end{equation}
where $\hat{\bm{s}}$ is the estimate of $\bm{s}$, $\mathcal{F}^{\dagger}$ is the inverse STFT, $\odot $ is the element-wise product, $\mathcal{M}$ is a DNN for estimating a T-F mask, and $\theta$ is the set of its parameters.
Obviously, the quality of output signals is determined by the parameters of DNN.

In this study, we focus on the training strategy for the DNN, $\mathcal{M}( \:\cdot\: ; \theta )$.
The conventional training strategies (CTT and NeTT) are explained in this section, whereas the proposed training strategy (NyTT) will be introduced in Section \ref{sec:proposed}.

%\vspace{-6pt}
\subsection{Clean-target Training (CTT)}

Most of the literature of DNN-based speech enhancement is based on Clean-target Training\cite{Wang_2018}, which utilizes the clean speech signals as the training target.
It minimizes the following prediction error between the estimated signal $\hat{\bm{s}}$ and the clean target $\bm{s}$,
%\vspace{-2pt}
\begin{equation}
\mathcal{L}^{\mbox{\scriptsize CTT}} = 
\frac{1}{M} \sum_{m=1}^M
\mathcal{D}( \hat{\bm{s}}_m, \bm{s}_m ),
%\vspace{-2pt}
\end{equation}
where $M$ is the minibatch-size, and 
$\mathcal{D}$ is a function that measures the difference between the input variables, such as the $\ell_2$ distance 
$\mathcal{D}( \bm{a},\bm{b} ) = \lVert \bm{a}-\bm{b} \rVert_2^2$.
Since $\hat{\bm{s}}_m$ is predicted from $m$th noisy observation $\bm{x}_m$ as in Eq.~\eqref{eq:DNNusage}, Clean-target Training requires pairs of noisy and clean signals $(\bm{x}_m,\bm{s}_m)$ as training data.
These training data are simulated by mixing clean speech signals and noise that are collected independently.
This is because recording such paired signals in the real environment is not feasible.

Clean-target Training has an essential limitation due to the requirement of clean speech signals.
In the real use cases, recording conditions have extreme variation caused by recording equipment, mouth-microphone angle and distance, surrounding environments, and several other factors.
Covering all possible recording conditions by studio-recorded speech signals is impossible because clean signals can only be acquired in the well-controlled environment in a studio. 
This fact limits the amount and variation of the training data, which may degrade the performance of speech enhancement.

%\vspace{-6pt}
\subsection{Noise-target Training (NeTT)}
\label{sec:NeTT}
Another training strategy, which was originally proposed for image processing~\cite{noise2noise}, is Noise-target Training\cite{n2n_speech}.
It trains a DNN to predict a noisy signal from another noisy signal as follows.
Noise-target Training considers two different noises $\bm{n}^{(1)}$ and $\bm{n}^{(2)}$ for a clean target $\bm{s}$.
Then, two kinds of observations can be obtained as $\bm{x}^{(1)} \!= \bm{s} + \bm{n}^{(1)}$ and
$\bm{x}^{(2)} \!= \bm{s} + \bm{n}^{(2)}$, which form a pair of noisy signals $(\bm{x}^{(1)},\bm{x}^{(2)})$.
Using such noisy-noisy pairs $(\bm{x}^{(1)}_m,\bm{x}^{(2)}_m)$, a DNN is trained to minimize the following prediction error between the output signal $\hat{\bm{s}}^{(1)}$ estimated from $\bm{x}^{(1)}$ and $\bm{x}^{(2)}$,
%\vspace{-2pt}
\begin{equation}
\mathcal{L}^{\mbox{\scriptsize NeTT}} = 
\frac{1}{M} \sum_{m=1}^M
\mathcal{D}( \hat{\bm{s}}_m^{(1)}, \bm{x}_m^{(2)} ).
\label{eq:L_NeTT}
%\vspace{-2pt}
\end{equation}
Since random noise cannot be predicted by a DNN, the random components contained in the training data are mapped to their expected values.
Therefore, by assuming the noise as zero-mean random variable, this training strategy yields a DNN that eliminates the noise.

Although Noise-target Training is useful for image processing~\cite{noise2noise}, it must inherit the limitation of Clean-target Training for speech enhancement.
For image applications, the above noisy-noisy pairs (contaminated by shot noise and thermal noise) can be easily obtained by a camera with short exposures.
In contrast, for audio applications, it is impossible to observe multiple noisy signals with exactly the same speech because audio signals are time- and space-variant.
Hence, we must use the clean speech signals to simulate the noisy-noisy pairs, which limits the variation of the training data.

%\vspace{-4pt}
\section{Proposed method}
\label{sec:proposed}
%\vspace{-2pt}

The success of Noise-target Training has suggested a possibility of training a DNN without clean signals.
However, it is not in a suitable form for audio applications as discussed above.
In this study, we investigate another possibility of training without clean signals, Noisy-target Training, that is suitable for speech enhancement.

%\vspace{-6pt}
\subsection{Noisy-target Training (NyTT)}
\label{sec:NyTT}

The above Noise-target Training has revealed two facts: (1) clean signals are not mandatory for training, and (2) noisy signals can be utilized instead.
By interpreting them in the broadest sense, we propose Noisy-target Training as follows.

% In the proposed training strategy, we only require noisy observation $\bm{x}$ and additional noise $\bm{z}$ which is a different noise from $\bm{n}$, i.e., clean signals are not utilized.
In the proposed training strategy, we only require noisy signal $\bm{x}$ and noise $\bm{n}$, i.e., clean signals are not utilized.
By mixing them, a \textit{more noisy signal} $\bm{y}$ is synthesized as follows:
%\vspace{-2pt}
\begin{equation}
\bm{y} = \bm{x} + \bm{n}
%\vspace{-2pt}
\end{equation}
This forms a pair of more noisy and noisy signals $(\bm{y},\bm{x})$.
%This forms a pair of noisy and more noisy signals $(\bm{y},\bm{x})$.
By inputting the more noisy signal $\bm{y}$ into the DNN as
%\vspace{-2pt}
\begin{equation}
\hat{\bm{s}} = 
\mathcal{F}^{\dagger} \left(
\mathcal{M}( \bm{y}; \theta ) \odot 
\mathcal{F} \left( \bm{y} \right) \right),
%\vspace{-2pt}
\end{equation}
the proposed method trains the DNN by minimizing the following prediction error between $\bm{x}$ and the enhanced more noisy signal $\hat{\bm{s}}$,
%\vspace{-2pt}
\begin{equation}
\mathcal{L}^{\mbox{\scriptsize NyTT}} = \frac{1}{M} \sum_{m=1}^M \mathcal{D}( \hat{\bm{s}}_m, \bm{x}_m).
%\vspace{-2pt}
\end{equation}
Therefore, the proposed method realizes training similar to Eq.~\eqref{eq:L_NeTT} without using any clean signal.

Since $\bm{x} = \bm{s} + \bm{n}^{(\mathrm{obs})}$, the proposed method can be viewed as Noise-target Training in Section \ref{sec:NeTT} with $\bm{x}^{(1)} \!= \bm{s}+\bm{n}^{(\mathrm{obs})}\!+\bm{n}$ and $\bm{x}^{(2)} \!= \bm{s}+\bm{n}^{(\mathrm{obs})}$.
Hence, the validity of the proposed method depends on the statistics of $\bm{n}^{(\mathrm{obs})}\!+\bm{n}$ and $\bm{n}^{(\mathrm{obs})}$ utilized during the training.
Since we suppose that both $\bm{n}^{(\mathrm{obs})}$ and $\bm{n}$ are given by real-world recordings, theoretical validation cannot be made for the proposed method.
Even so, our experiments in the next section confirmed its effectiveness for speech enhancement.

%\vspace{-4pt}
\section{Experiments}
\label{sec:experiments}
%\vspace{-2pt}

%-----------------------------------------------------------------
We conducted three types of experiments to investigate the performance of the proposed method (NyTT):
%\vspace{-3pt}
\begin{description}[leftmargin=20pt]
  \setlength{\parskip}{0cm}
  \setlength{\itemsep}{0cm}
\item[Proof of concept:] We investigated whether NyTT can train the DNN without clean speech. We evaluated the performance on both seen and unseen datasets, i.e., w/ and w/o mismatch between training and testing data.
%We investigated whether NyTT properly enables training of a DNN without clean speech. We evaluated the performance on both seen and unseen datasets, i.e., w/ and w/o mismatch between training and testing data.
\item[Effects of SNR of noisy target:] We investigated the performance of NyTT in response to the SNR of noisy target.
%We investigated the performance of NyTT in response to the SNR between $\bm{s}$ and $\bm{n}^{(\mathrm{obs})}$.
\item[Effects of types of additional noise:] We investigated the performance of NyTT in response to the relationship between $\bm{n}^{(\mathrm{obs})}$ and $\bm{n}$ by using four types of additional noise datasets.
\end{description}
%\vspace{-3pt}
%-----------------------------------------------------------------

%\vspace{-6pt}
\subsection{Experimental setups}
\label{sec:setups}

\textbf{Datasets:} Table\,\ref{tbl:noisy_dataset} shows datasets used in experiments.
We utilized the {\sf VoiceBank-DEMAND}~\cite{voicebankdemand} which is openly available and frequently used in the literature of DNN-based speech enhancement~\cite{Koizumi_icassp_2020, kawanaka}. 
%\cite{Pascual_2017, Koizumi_icassp_2020, kawanaka}.
The train and test sets consists of 28 and 2 speakers (11572 and 824 utterances), respectively.
In addition to this dataset, to evaluate the performance under a training/testing data mismatched condition, we constructed a test dataset by mixing TIMIT~\cite{timit} (speech) and TAU Urban Acoustic Scenes 2019 Mobile~\cite{tau-2019} (noise) as {\sf TIMIT-MOBILE} at signal-to-noise ratio (SNR) randomly selected from $-5$, $0$, $5$, and $10$~dB.
The test sets consist of 1680 utterances spoken by 168 speakers (112 males and 56 females).

To mimic the use of noisy signals for training in NyTT, we additionally used {\sf Libri-Task1} and {\sf CHiME5} as noisy datasets.
{\sf Libri-Task1} consists of mixed signals of the development sets of LibriTTS~\cite{libritts} and TAU Urban Acoustic Scenes 2020 Mobile~\cite{tau-2020} (TAU-2020) whose SNR was randomly selected from $0$, $5$, $10$, and $15$~dB. This dataset includes 8.97 hours of noisy speech with 5736 utterances.
{\sf CHiME5} was the training dataset of the 5th CHiME Speech Separation and Recognition Challenge~\cite{chime5}, and consisted of 77.24 hours of noisy speech with 79967 utterances which was created by cutting each speech interval in the continuous training data with before/after 0.5 sec margin.
In addition, we used background noise of CHiME3~\cite{chime3} as noise dataset (CHiME3).

\begin{table}[tt]
%\vspace{-2pt}
\caption{List of training/testing datasets. {\sf Libri-Task1} and {\sf CHiME5} include only pre-mixed (noisy) signals, which imitates the situation that only noisy speech signals $\bm{x}$ are available.}
\label{tbl:noisy_dataset}
\begin{center}
\vspace{-10pt}   
\small 
\begin{tabular}{l|ll} 
%\hline
\toprule
Name & Clean $\bm{s}$ & Noise $\bm{n}^{(\mathrm{obs})}$\\
\midrule
{\sf VoiceBank-DEMAND}~\cite{voicebankdemand} & VoiceBank~\cite{voicebank} & DEMAND~\cite{demand}\\
{\sf TIMIT-MOBILE} & TIMIT~\cite{timit} & TAU-2019~\cite{tau-2019}\\
\midrule
{\sf Libri-Task1} & \multicolumn{2}{c}{Libri-TTS~\cite{libritts} + TAU-2020~\cite{tau-2020}}\\
{\sf CHiME5} & \multicolumn{2}{c}{Only noisy signal provided}\\
\bottomrule
  \end{tabular}
\end{center}
\vspace{-15pt}
\end{table}

%-----------------------------------------------------------------
%\vspace{5pt}
\noindent
\textbf{Comparison methods and metrics:} 
In order to investigate whether NyTT can solve the recording condition mismatch problem by utilizing a larger amount of noisy target, we evaluated the following two versions of NyTT.
These methods were compared with CTT and NeTT~\cite{n2n_speech}.

As the metrics, we used CSIG, CBAK, COVL~\cite{Hu_2008}, PESQ, and scale-invariant signal-to-distortion ratio (SI-SDR). The first four metrics are the standard metrics used in {\sf VoiceBank-DEMAND}, and SI-SDR is a metrics widely used for evaluation of speech enhancement.
%CSIG, CBAK, and COVL are the popular predictor of the mean opinion score (MOS) of the target signal distortion, background noise interference, and overall speech quality, respectively~\cite{}.

%-----------------------------------------------------------------
%\vspace{5pt}
\noindent
\textbf{Training details:}
For NyTT and NyTT~(L), we randomly selected an additional noise $\bm{n}$ from DEMAND, TAU-2020, and CHiME3, and mixed to noisy speech $\bm{x}$ at randomly selected SNRs between $-5$ to $5$~dB, where SNR is measured by considering $\bm{x}$ as the signal and $\bm{n}$ as noise. 
For CTT, to use the same variety of noise samples as NyTT, we augmented the noisy dataset by randomly swapping the noise in noisy signals to a noise in DEMAND, TAU-2020, and CHiME3.
In the same sense, we used DEMAND, TAU-2020, and CHiME3 as noise dataset for NeTT.
Therefore, the amount and variety of training noise samples were the same for all methods, and only the type and amount of the target signals were different.

The DNN estimated a complex-valued T-F mask%~\cite{williamson_2016}
, and consisted of a CNN-BLSTM which has the same architecture of \cite{kawanaka}.
The input of the DNN was log-amplitude spectrogram of the input signal.
%The input of the DNN was log-amplitude spectrogram of the input signal whose size was $F \times K$.
The spectrogram of the input was multiplied by the estimated complex T-F mask and transformed back to the time-domain, where the STFT parameters, frame shift and window size $(=\text{DFT size})$, were set to 128- and 512-samples, respectively, with the Hamming window.

We used mean-squared-error (MSE) as $\mathcal{D}( \bm{a},\bm{b} ) = \frac{1}{T} \lVert \bm{a}-\bm{b} \rVert_2^2$ and Adam optimizer with a fixed learning rate 0.0001.%~\cite{adam}
We separated the training dataset into randomly selected 50 and other utterances, and used as validation and development datasets, respectively.
We trained DNNs 500 epochs with batchsize 50, and finally used the model with the best validation SI-SDR.
In CTT and NeTT, noise $\bm{n}$ was added to $\bm{s}$ at SNR randomly selected from $-5$, $0$, $5$, and $10$~dB.

\vspace{-5pt}
\subsection{Proof of concept}
\label{sec:objective}

% %%%%%%%%%%%%%%%%%%%%%%%%%%%%%%%%%%%%%%%%%%%%%%%%%%%%%%%%
% \begin{table}[tt]
% %\vspace{-10pt}  
% \caption{List of training/testing datasets. {\sf Libri-Task1} and {\sf CHiME5} include only pre-mixed (noisy) signals, which imitates the situation that only noisy speech signals $\bm{x}$ are available.}
% \label{tbl:noisy_dataset}
% \begin{center}
% \vspace{-6pt}   
% \small 
% \begin{tabular}{l|ll} 
% %\hline
% \toprule
% Name & Clean $\bm{s}$ & Noise $\bm{n}^{(\mathrm{obs})}$\\
% \midrule
% {\sf VoiceBank-DEMAND}~\cite{voicebankdemand} & VoiceBank~\cite{voicebank} & DEMAND~\cite{demand}\\
% {\sf TIMIT-MOBILE} & TIMIT~\cite{} & TAU-2019~\cite{}\\
% \midrule
% {\sf Libri-Task1} & \multicolumn{2}{c}{Libri-TTS~\cite{} + TAU-2020~\cite{}}\\
% {\sf CHiME5} & \multicolumn{2}{c}{Only noisy signal provided}\\
% \bottomrule
%   \end{tabular}
% \end{center}
% \vspace{-10pt}  
% \end{table}
% %%%%%%%%%%%%%%%%%%%%%%%%%%%%%%%%%%%%%%%%%%%%%%%%%%%%%%%%

\begin{table}[tt]
\caption{Results on {\sf VoiceBank-DEMAND} (no mismatch of training and testing datasets).
Input means the scores of input signals.
}
\label{tbl:voicebank}
\begin{center}
\vspace{-5pt}
\small 
\begin{tabular}{l|ccccc} 
\toprule
    Method & SI-SDR & PESQ & CSIG & CBAK & COVL \\ \midrule
    Input
    & 9.21 & 1.97 & 3.35 & 2.44 & 2.63 \\ 
    CTT
    & {\bf 19.53} & {\bf 2.68} & {\bf 3.83} & {\bf 3.37} & {\bf 3.25} \\
    NeTT
    & 19.50 & 2.63 & 3.77 & 3.34 & 3.19 \\
    NyTT
    & 17.66 & 2.30 & 3.19 & 3.01 & 2.72\\
    NyTT (L)
    & 17.72 & 2.31 & 3.23 & 3.02 & 2.75 \\
\bottomrule
  \end{tabular}
\vspace{5pt}
\end{center}
\caption{Results on {\sf TIMIT-MOBILE} (with mismatch of training and testing datasets). Input means the scores of input signals.}
\label{tbl:timit}
\begin{center}
\vspace{-10pt}
\small 
\begin{tabular}{l|ccccc} 
\toprule
    Method & SI-SDR & PESQ & CSIG & CBAK & COVL \\ \midrule
    Input
    & 4.69 & 1.30 & 2.73 & 1.75 & 1.94 \\ 
    CTT
    & {\bf 12.60} & {\bf 2.02} & 3.22 & {\bf 2.71} & 2.58 \\
    NeTT
    & 12.26 & 1.99 & 3.13 & 2.67 & 2.52 \\
    NyTT
    & 12.09 & 1.95 & 3.41 & 2.61 & {\bf 2.64} \\
    NyTT (L)
    & 12.38 & 1.91 & {\bf 3.43} & 2.58 & 2.63 \\
\bottomrule
  \end{tabular}
\end{center}
  \vspace{-10pt}
\end{table}

For the proof of concept, we conducted two experiments.
First, we conducted an experiment for verifying whether NyTT can train the DNN without clean speech signals. To remove the effect of training/testing data mismatch effect, we used the test dataset of {\sf VoiceBank-DEMAND} for evaluating scores. Table~\ref{tbl:voicebank} summarizes the evaluated scores, where NyTT and NyTT~(L) achieved higher scores than that of the input signal (Noisy).
This result indicates that Noisy-target Training can train a DNN without clean speech.

Next, we evaluated each method on {\sf TIMIT-MOBILE} to confirm whether NyTT robustly worked on unseen test data.
Here, there is a mismatch between recording conditions of training and testing datasets.
Table~\ref{tbl:timit} summarizes the evaluated scores.
While CTT and NeTT performed better than NyTT in the {\sf VoiceBank-DEMAND} results (no mismatch between testing and trainig datasets), the performance of all methods was similar in the results on {\sf TIMIT-MOBILE} (having mismatch between training and testing datasets).
Even though NyTT did not use any clean speech in training, it achieved results similar to those obtained by CTT and NeTT which used clean speech in training.
This result might indicate that training using clean speech can overfit to the signals in the training dataset.
The proposed NyTT has potential to avoid such overfitting by using a huge amount of noisy data which can be easily acquired.

\begin{figure}[t]
  \centering
  \centerline{\includegraphics[bb=0 0 720 302,width=\linewidth]{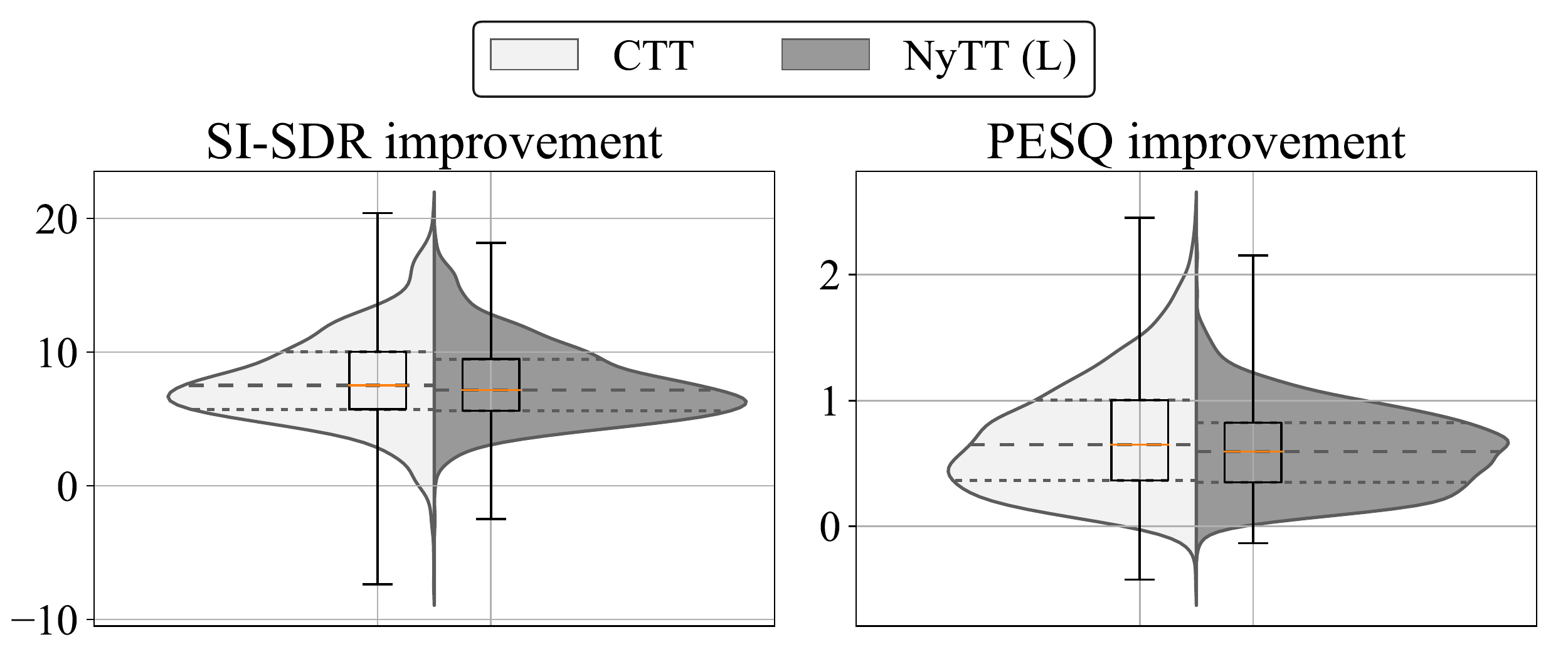}}
    \vspace{-5pt}
\caption{Comparison of SI-SDR/PESQ improvements between CTT and NyTT~(L) on {\sf TIMIT-MOBILE} (having mismatch with training data).}
\label{fig:violin}
  \vspace{-14pt}
\end{figure}

Fig.~\ref{fig:violin} shows SI-SDR/PESQ improvements of CTT and NyTT~(L) corresponding to Table~\ref{tbl:timit}.
The median of both training methods was almost the same, whereas the variance of NyTT was smaller than that of CTT. 
This suggests that NyTT has stable performance even when there is a mismatch between training and testing datasets.

%XXXXXX NEED TO CHANGE AFTER FIGURE REVISION XXXXXXXX.
%Figure ~\ref{fig:violin}では，最も高い評価はCTTで得られているが，中央値でNyTT(L)がCTTを上回っている．このことから，Noisy-target Trainingは、training/testing data mismatch がある状況でも、安定した性能が発揮できることが示唆された．

\vspace{-3pt}
\subsection{Effects of SNR of noisy target signal in NyTT}
\label{sec:require_noisy}

Since NyTT becomes CTT when SNR of the noisy target signal $\bm{x}$ is $\infty$ (i.e., $\bm{x}$ is clean), we investigated the relationship between SNR of noisy target and the performance of NyTT.
We modified {\sf VoiceBank-DEMAND} so that all noisy targets' SNR became $-5$, $0$, $5$, $10$, $15$ and $20$~dB.
The additional noise $\bm{n}$ was taken from DEMAND.
To remove the training/testing data mismatch effect, we used the test dataset of {\sf VoiceBank-DEMAND} for evaluation.

\begin{figure}[t]
  \centering
  \centerline{\includegraphics[width=\linewidth,bb=0 0 461 209]{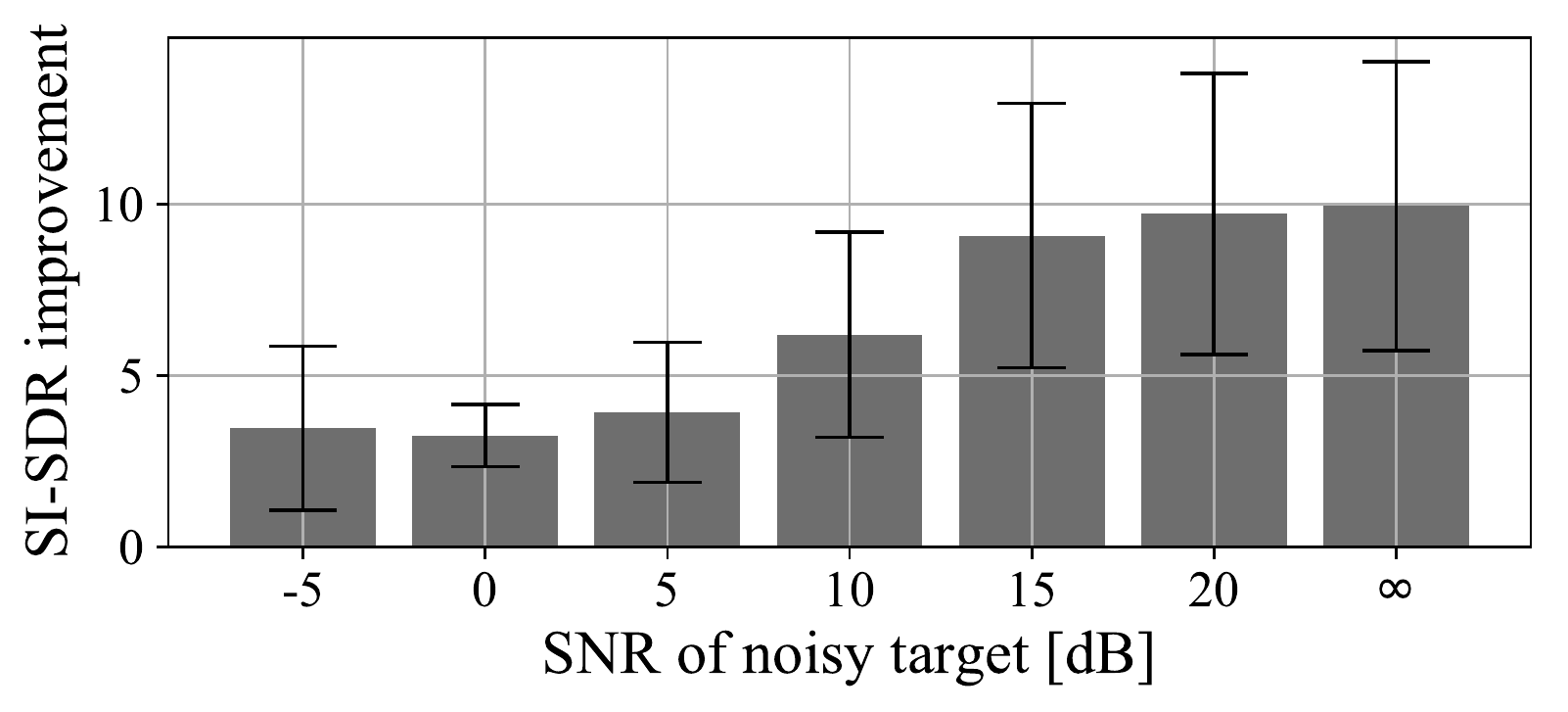}}
    \vspace{-5pt}
\caption{Relationship between SNR of noisy signal $\bm{x}$ utilized in NyTT and its SI-SDR improvement. 
SNR $\infty$ dB is equivalent to CTT.}
\label{fig:snr}
  \vspace{-10pt}
\end{figure}

Fig.~\ref{fig:snr} shows SI-SDR improvements for each SNR.
As in the figure, when SNR of the noisy target $\bm{x}$ was greater than $5$ dB, the performance increased as the SNR of $\bm{x}$ increased.
Meanwhile, there was almost no difference in SI-SDR improvement for $-5$ and $0$ dB SNR conditions.
This might be because when the power of $\bm{n}^{(\mathrm{obs})}$ is equal to or greater than that of $\bm{s}$, a DNN is trained to predict not only $\bm{s}$ but also $\bm{n}^{(\mathrm{obs})}$ by removing only additional noise $\bm{n}$.
In contrast, when SNR of noisy target $\bm{x}$ was $15$ and $20$ dB, the performance was almost the same as CTT.
These results suggest that  (1) SNR of the noisy target $\bm{x}$ should be greater than $0$ dB for NyTT to be work, and (2) noisy speech signals with SNR greater than or equal to $15$ dB can serve as ``clean'' signals for training in speech enhancement.

\vspace{-2pt}
\subsection{Effects of types of additional noise used in NyTT}
\label{sec:require_noise}

Since NyTT utilized noisy signals $\bm{x} = \bm{s} + \bm{n}^{(\mathrm{obs})}$ with additional noise $\bm{n}$, we investigated the relation of the types of noise used for $\bm{n}^{(\mathrm{obs})}$ and $\bm{n}$.
For the noisy signals $\bm{x}$, {\sf VoiceBank-DEMAND} was utilized as in the previous experiments, i.e., $\bm{n}^{(\mathrm{obs})}$ is from DEMAND.
For the additional noise $\bm{n}$, we utilized one of the following four datasets: DEMAND, TAU-2020, CHiME3, and the training dataset of DCASE2016 Challenge Task 2 (Task2)\cite{dcase2016}.
These four noise datasets can be classifies into two: the first three datasets include various environmental noise, while Task2 includes only monophonic sound events that would occur in an office.
For the testing dataset, {\sf TIMIT-NOISEX-92} was utilized to avoid using TAU-2019 which is similar to TAU-2020.
It was generated by mixing TIMIT (speech) and NOISEX-92~\cite{noisex92} (noise) at SNR randomly selected from $0$, $5$, $10$, and $15$~dB.
Note that the type of noise in NOISEX-92 is different from all four datasets used for the additional noise $\bm{n}$.

% We investigated how effect the characteristics of the additional noise for Noisy-target Training.
% We mixed noise signal from four different noise datasets into a noisy signal in the {\sf VoiceBank-DEMAND}, and evaluated the results when each was used as more-noisy input for Noisy-target Training.
% As the additional noise datasets, we used DEMAND, TAU-2020, CHiME3, and the training dataset of DCASE2017 Challenge Task 2 (Task2).
% In contrast to the first three dataset includes various environmental noise, Task2 includes only monophonic sound events that would occur in an office.
% Since TAU-2020 is similar to TAU-2019, we used {\sf TIMIT-NOISEX-92} as the testing dataset, which is generated by mixing TIMIT (speech) and NOISEX-92~\cite{noisex92} (noise) at a randomly selected from 0, 5, 10, 15~dB SNR whose noise is different from all four additional noise datasets.

\begin{table}[tt]
\vspace{-7pt}
\caption{Comparison on type of additional noise $\bm{n}$ in NyTT.}
\label{tbl:noise}
\begin{center}
\vspace{-12pt}
\small 
\begin{tabular}{l|ccccc} 
%\hline
\toprule
     & Input & DEMAND & TAU-2020 & CHiME3 & Task2 \\ %\hline
\midrule
    SI-SDR & 8.47 & 12.09 & 12.00 & 11.99 & 9.63 \\ 
    PESQ & 1.44 & 1.74 & 1.83 & 1.95 & 1.52 \\
    %\hline
\bottomrule
  \end{tabular}
\end{center}
\vspace{-12pt}
\end{table}

\begin{figure}[t]
  \centering
  \centerline{\includegraphics[width=\linewidth,bb=0 0 720 360]{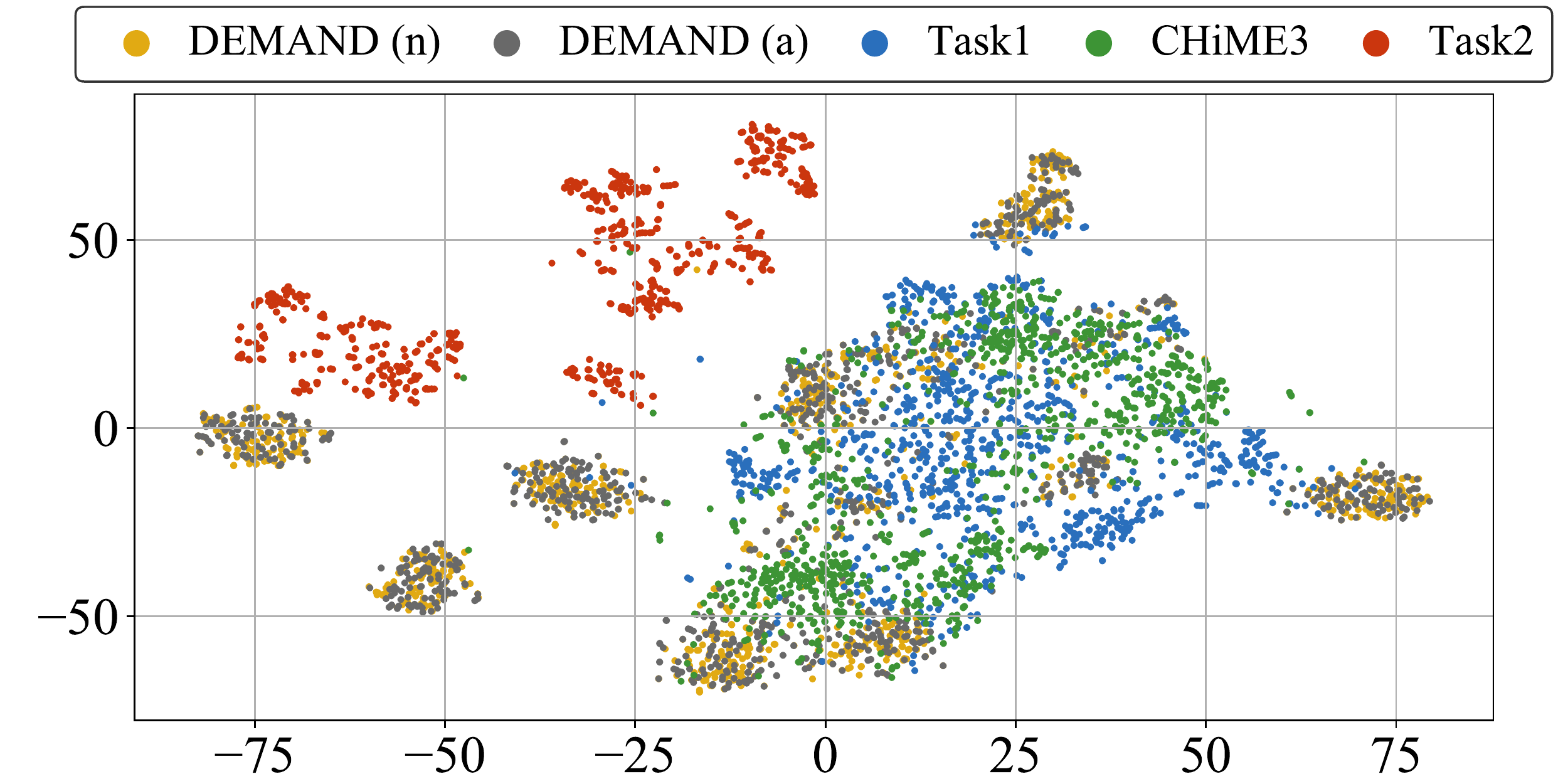}}
  \vspace{-2pt}
  \caption{Visualization of distribution of $\bm{n}$ and $\bm{n}^{(\mathrm{obs})}$ using t-SNE.
  DEMAND(n) denotes $\bm{n}^{(\mathrm{obs})}$, 
  and the others are $\bm{n}$ in Table~\ref{tbl:noise}.
  }
  \label{fig:tSNE}
  \vspace{-12pt}
\end{figure}

Table~\ref{tbl:noise} shows SI-SDR and PESQ of NyTT using one of the four datasets for $\bm{n}$.
NyTT using DEMAND, TAU-2020, CHiME3 achieved similar scores, whereas NyTT with Task2 failed to enhance the signal (the scores of Input and Task2 were almost the same).
This should be because Task2 contains a very different type of noise as mentioned in the previous paragraph.
To confirm it, we visualized distribution of the datasets as shown in Fig.~\ref{fig:tSNE}.
This figure was obtained as follows.
First, we randomly selected 1000 samples from each training dataset and extracted the first 2 sec to align the length of data.
Second, we calculated the acoustic feature of each sample using VGGish~\cite{vggish}.
Finally, the calculated features were illustrated as a 2D map by t-Distributed Stochastic Neighbor Embedding (t-SNE)~\cite{tsne}.
From the figure, it can be seen that DEMAND, TAU-2020 and CHiME3 are similarly distributed, but Task2 has almost no overlap with them.
This result suggests that NyTT can successfully train a DNN when the distribution of additional noise $\bm{n}$ can hide in the distribution of $\bm{n}^{(\mathrm{obs})}$.
This should be because the different type of noise $\bm{n}$ (as Task2 in this experiment) can be distinguished from the noisy signals $\bm{x} = \bm{s} + \bm{n}^{(\mathrm{obs})}$, which enables a DNN to eliminate only the additional noise $\bm{n}$ while keeping $\bm{n}^{(\mathrm{obs})}$.

%$\bm{s}$ and $\bm{n}^{(\mathrm{obs})}$

%\vspace{-2pt}
\section{Conclusions}
\label{sec:con}
%\vspace{-2pt}
In this study, for DNN-based speech enhancement, we proposed a training strategy that does not require clean signals.
We utilized noisy signals as the target and trained a DNN to predict them from the more noisy signals (Section~\ref{sec:NyTT}).
Our experiments showed that the proposed method (1) was able to train a DNN without clean speech signals, (2) achieved the results similar to those obtained by using clean signals as the target when the training and testing datasets have a mismatch, and (3) revealed the borderline (15 dB) where a signal can be treated as clean in the training.
Future work includes evaluation using a larger dataset and theoretical validation.

%今後の課題・数学的な証明・ネットワークの表現能力が非常に高い場合は今回の手法は適用できない可能性があるため，ネットワークによる性能の違いを調査すべきである．・出力音声にはノイズが残りやすい傾向がある．

\end{document}